\documentclass[11pt]{article}
\usepackage{moriond,epsfig}

\def\Journal#1#2#3#4{{#1} {\bf #2}, #3 (#4)}

\def\NCA{{\em Nuovo Cimento} A}

\def\PLB{{\em Phys. Lett.}  B}
\def\PRL{\em Phys. Rev. Lett.}
\def\PRD{{\em Phys. Rev.} D}
\def\PRC{{\em Phys. Rev.} C}

\def\PDG{{\em J. Phys.} G}
\def\EPJ{{\em Eur. Phys. J.} C}
\def\RMP{\em Rev. Mod. Phys.}

\def\be{\begin{equation}}
\def\ee{\end{equation}}
\def\bea{\begin{eqnarray}}
\def\eea{\end{eqnarray}}

\begin{document}
\vspace*{4cm}
\title{Light Hadrons and New Enhancements in J/$\psi$ Decays at BESII}

\author{Guofa XU
        \\(for the BES Collaboration)}

\address{Institute of High Energy Physics, CAS, Beijing 100049, China}

\maketitle \abstracts{Based on 58 million $J/\psi$ samples collected
by the BESII detector at the BEPC, many mesons, baryons, and new
resonances have been reported. Here, I will review some recent
results of glueball candidates and new enhancement.}

\section{Introduction}
In this paper, some recent BESII results are reported based on 58
million $J/\psi$ events collected by the BESII detector at the BEPC.
For much more detail, please see the references.

\section{Scalars ($0^{++}$)}
As we know that so many scalars are listed in PDG06~\cite{pdg06},
but according to the quark model no enough room for all of these
scalar particles. On the other hand, the Lattice QCD predicted that
the ground state glueball is $0^{++}$, and its mass is around
1.5$\sim$1.8 GeV. Theoretical physicists expect that glueballs will
mix with nearby $q\bar{q}$ states of the same quantum
numbers~\cite{glue1,glue2}, it makes the situation more difficult
for the glueball identification. Although the identification of a
glueball is very complicated, there are several glueball candidates,
such as $f_0(1500)$ and $f_0(1710)$, considering the possible mix
with the ordinary $q\bar{q}$ meson, $f_0(1370)$, $f_0(1500)$,
$f_0(1710)$, and $f_0(1790)$ have been analyzed for more detail by
using the partial wave analyzes (PWA) method in $J/\psi\to \gamma
\pi\pi$, $\gamma K\bar{K}$, $J/\psi\to \omega K\bar{K}$, and
$J/\psi\to \phi\pi\pi$, $\phi K\bar{K}$ channels.

\subsection{The Analysis of $J/\psi\to \gamma\pi\pi$ and $\gamma K\bar{K}$ Channels}
\begin{figure}[hbt]
\begin{minipage}{65mm}
\centerline{
    \psfig{file=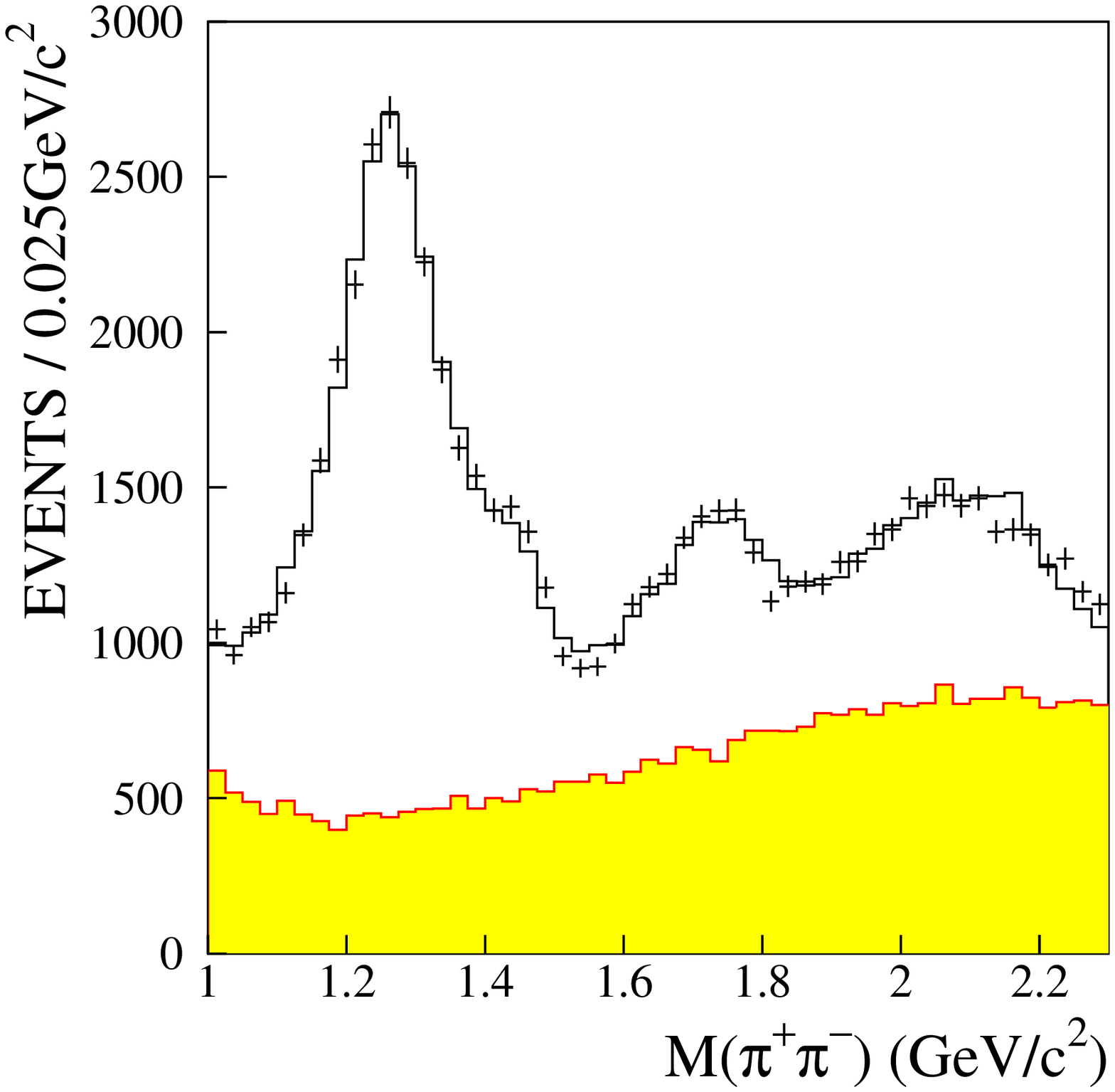,width=1.0\textwidth}}
    \caption{The $\pi^+\pi^-$ invariant mass distribution from
$J/\psi \to \gamma \pi^+\pi^-$. The crosses are data, the full
histogram shows the maximum likelihood fit, and the shaded histogram
corresponds to the $\pi^+\pi^-\pi^0$ background.}
    \label{gpipic}
\end{minipage}
\hspace{\fill}
\begin{minipage}{65mm}
\centerline{
    \psfig{file=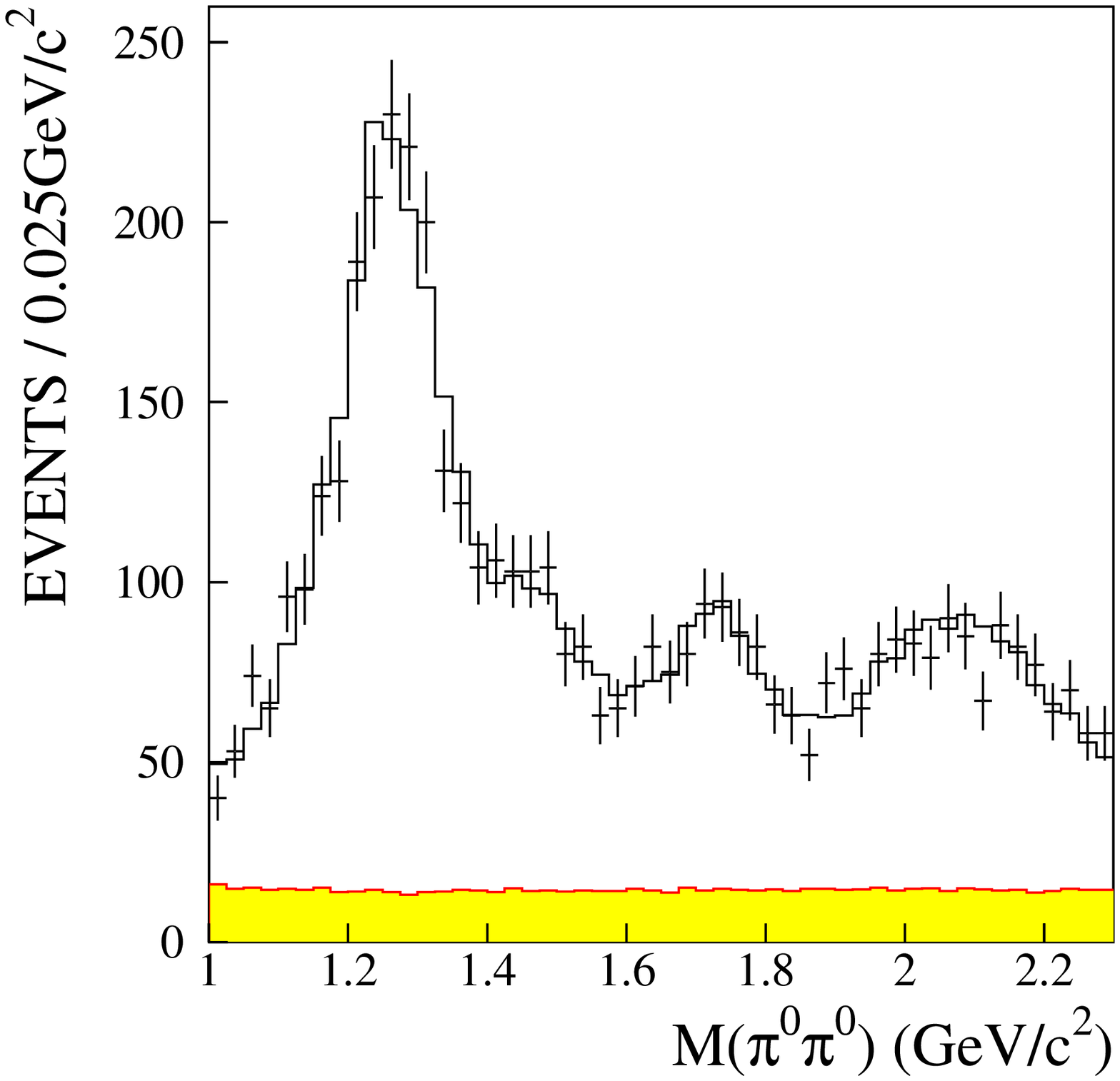,width=1.0\textwidth}}
    \caption{The $\pi^0\pi^0$ invariant mass distribution from
$J/\psi \to \gamma \pi^0\pi^0$. The crosses are data, the full
histogram shows the maximum likelihood fit, and the shaded histogram
corresponds to the background.}
    \label{gpipi0}
\end{minipage}
\end{figure}

  The partial wave analyzes of $J/\psi\to\gamma\pi^+\pi^-$
  and $J/\psi\to\gamma\pi^0\pi^0$ show the evidence for two $0^{++}$ states around
  the 1.45 and 1.75 GeV/$c^2$ mass regions (Fig.~\ref{gpipic},~\ref{gpipi0})~\cite{gpipi}.
  The $f_0(1500)$ has a mass of $1466\pm
  6\pm 20$ MeV/$c^2$, a width of $108{^{+14}_{-11}}\pm 25$ MeV/$c^2$, and
  a branching fraction
  ${\cal B}(J/\psi\to \gamma f_0(1500)\to\gamma \pi^+\pi^-) =
  (0.67\pm0.02\pm0.30)\times 10^{-4}$.
  The $0^{++}$ state in the $\sim$1.75 GeV/$c^2$ mass region has a mass of
  $1765{^{+4}_{-3}}\pm 13$ MeV/$c^2$ and a width of $145\pm 8 \pm 69$ MeV/$c^2$.

\begin{figure}[hbt]
\centerline{
    \psfig{file=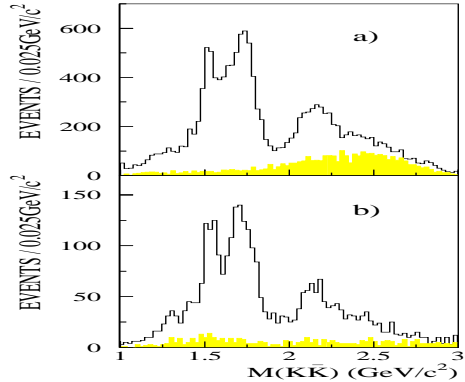,width=60mm,height=5cm}}
\caption{Invariant mass spectra of a) $K^+K^-$, b) $K^0_SK^0_S$ for
$J/\psi \to \gamma K \bar K$ events, where the shaded histograms
correspond to the estimated background contributions.}
    \label{gkk}
\end{figure}

\noindent
 The PWA of $J/\psi\to\gamma K^+K^-$ and $J/\psi\to\gamma
K^0_SK^0_S$ show strong production of the $f'_2(1525)$ and the
S-wave resonance $f_0(1710)$ (Fig.~\ref{gkk})~\cite{gamkk}. The
$f_0(1710)$ peaks at a mass of $1740\pm 4^{+10}_{-25}$ MeV with a
width of $166{^{+5}_{-8}}{^{+15}_{-10}}$ MeV.
\par

\subsection{The Analysis of $J/\psi\to \omega K^+ K^-$ Channel}

\begin{figure}[htb]
\centerline{
  \epsfig{file=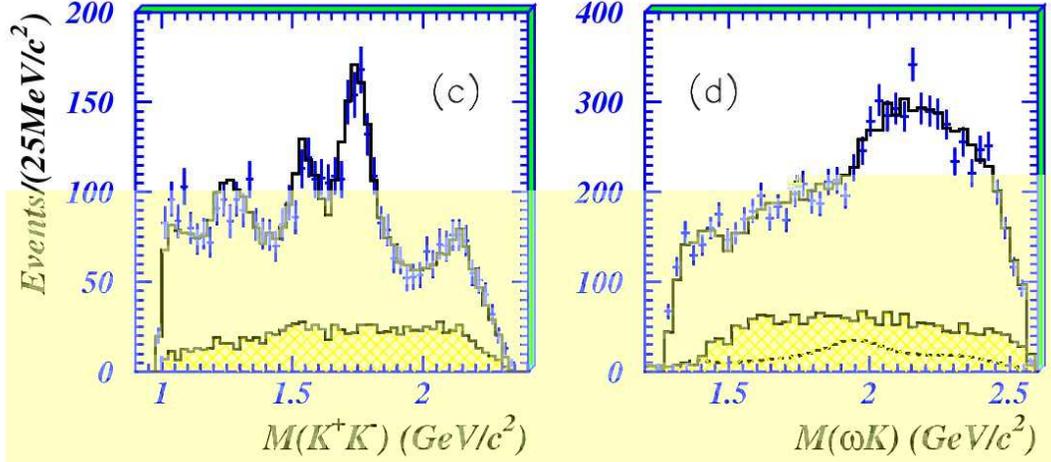,width=14.0cm}}
  \caption{(c) and (d) are projections on to $K^+K^-$ and $\omega K$ mass.
  Histograms show the maximum likelihood fit; the shaded region
indicates the background estimated from sidebins; the dashed curve
in (d) shows the magnitude of the $K_1(1400)$ contribution and a
$K\omega$ contribution at 1945 MeV/c$^2$.}
   \label{wkk}
\end{figure}

From Fig.~\ref{wkk}, one can see that a dominant feature is
$f_0(1710)$~\cite{omekk}. The fitted $f_0(1710)$ optimizes at $M =
1738 \pm 30$ MeV$/c^2$, $\Gamma = 125 \pm 20$ MeV$/c^2$.

\subsection{The Analysis of $J/\psi\to \phi\pi^+ \pi^-$ and $\phi K^+ K^-$ Channels}
After the partial wave analyzes for these $\phi\pi\pi$ and $\phi KK$
channels~\cite{phikk}, the data reported here have three important
features. Firstly, the parameters of $f_0(980)$ are all well
determined. Secondly, there is the clearest signal to date of
$f_0(1370) \to \pi ^+\pi ^-$; a resonant phase variation is
required, from interference with $f_2(1270)$. Thirdly, there is a
clear peak in $\pi \pi$ at 1775 MeV/c$^2$, consistent with
$f_0(1790)$; spin 2 is less likely than spin 0.

\begin{figure}[htb]
\centerline{\epsfig{file=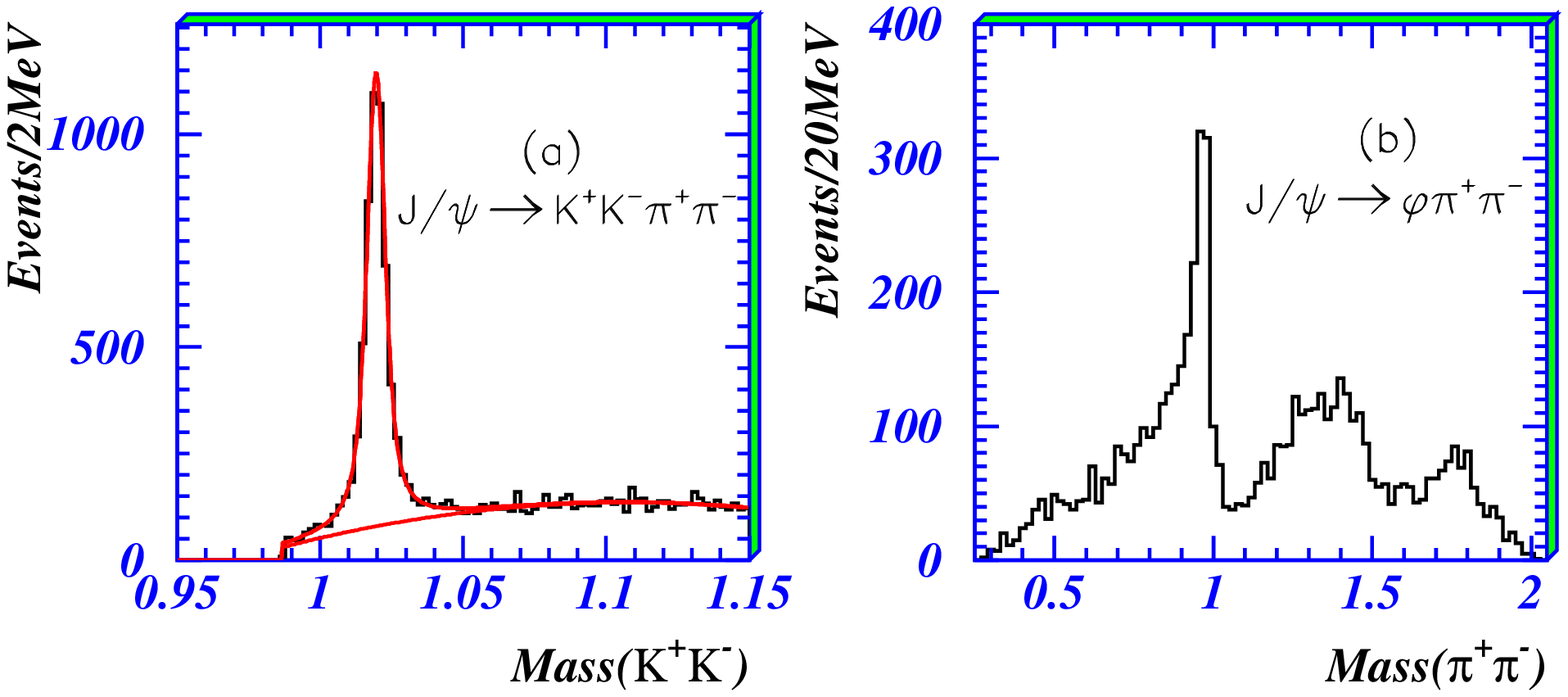,width=5.0in,height=3cm}}
\centerline{\epsfig{file=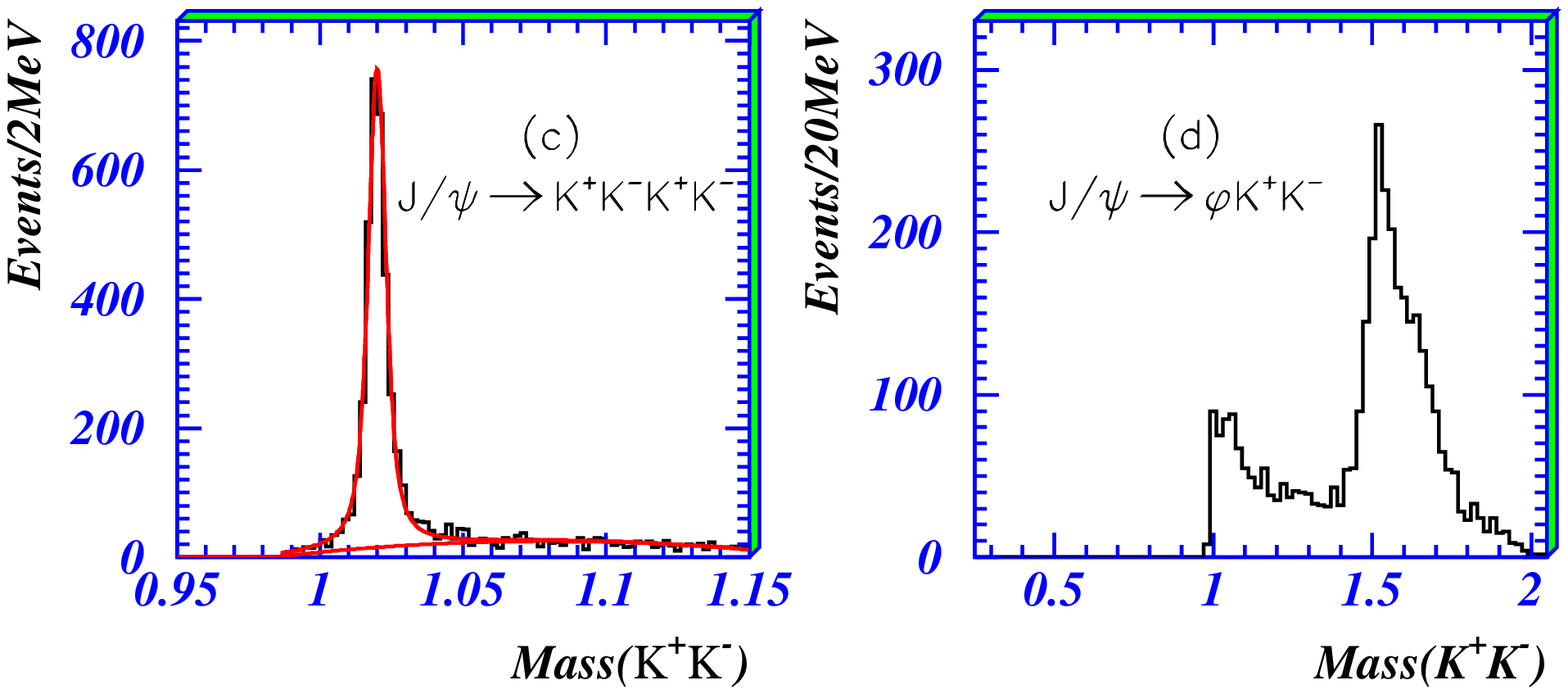,width=5.0in,height=3cm}}
\caption{The $K^+K^-$ invariant mass distributions for (a) $J/\psi
\to K^+K^-\pi ^+\pi ^-$, (c) $J/\psi \to K^+K^-K^+K^-$; curves show
the fitted background and a Gaussian fit to the $\phi$; (b) and (d)
show mass projections for events selected within $\pm 15$ MeV/c$^2$
of the $\phi$. } \label{phikpi}
\end{figure}

\par
\noindent
 In summary, (1) $f_0(1370)$ has been seen in $J/\psi\to
\phi\pi\pi$, but not in $J/\psi\to \omega\pi\pi$. (2) No peak of the
$f_0(1500)$ directly seen in $J/\psi\to\phi KK$, $\omega KK$,
$\phi\pi\pi$, and $\omega\pi\pi$, but in proton-proton scattering is
quite clear. (3) $f_0(1710)$ is observed clearly in both $J/\psi\to
\phi KK$ and $J/\psi\to\omega KK$, but with $Br(J/\psi\to\omega
f_0(1710)\to\omega KK)/Br(J/\psi\to\phi f_0(1710)\to \phi KK)\sim
6$, which is against a simple $s\bar{s}$ configuration for this
state. (4) $f_0(1790)$ which is seen in $\pi\pi$ rather than
$K\bar{K}$.
\par
\noindent Different models have different interpretations for these
experimental results. One of the interpretations is from
Cheng~\cite{cheng}, he explained that (1) $f_0(1710)$ is composed
primarily of the scalar glueball. (2) $f_0(1500)$ is close to an
$SU(3)$ octet. The glueball content of $f_0(1500)$ is very tiny
because an $SU(3)$ octet does not mix with the scalar glueball. (3)
$f_0(1370)$ consists of an approximate $SU(3)$ singlet with some
glueball component ($\sim 10\%$).

\section{Pesudo-scalars ($0^{-+}$)}
The first observation of $\eta(1440)$ was made in $p\bar{p}$
annihilation at rest into $\eta(1440)\pi^+\pi^-$, $\eta(1440)\to
K\bar{K}\pi$~\cite{bail}. Nowadays, The existence of two overlapping
pseudo-scalar states has been suggested to instead of the
$\eta(1440)$: one around 1405 MeV/c$^2$ decays mainly through
$a_0(980)\pi$ (or direct $K\bar{K}\pi$), and the other around 1475
MeV/c$^2$ mainly to $ K^*(892)\bar{K}$~\cite{pdg06,godfrey}. It is
therefore conceivable that the higher mass state is the $s\bar{s}$
member of the $2^1S_0$ nonet~\cite{rath,rath1}, while the lower mass
state may contain a large gluonic content~\cite{close,close1}.
\par

\begin{figure}[hbt]
\begin{minipage}{65mm}
\centerline{
    \psfig{file=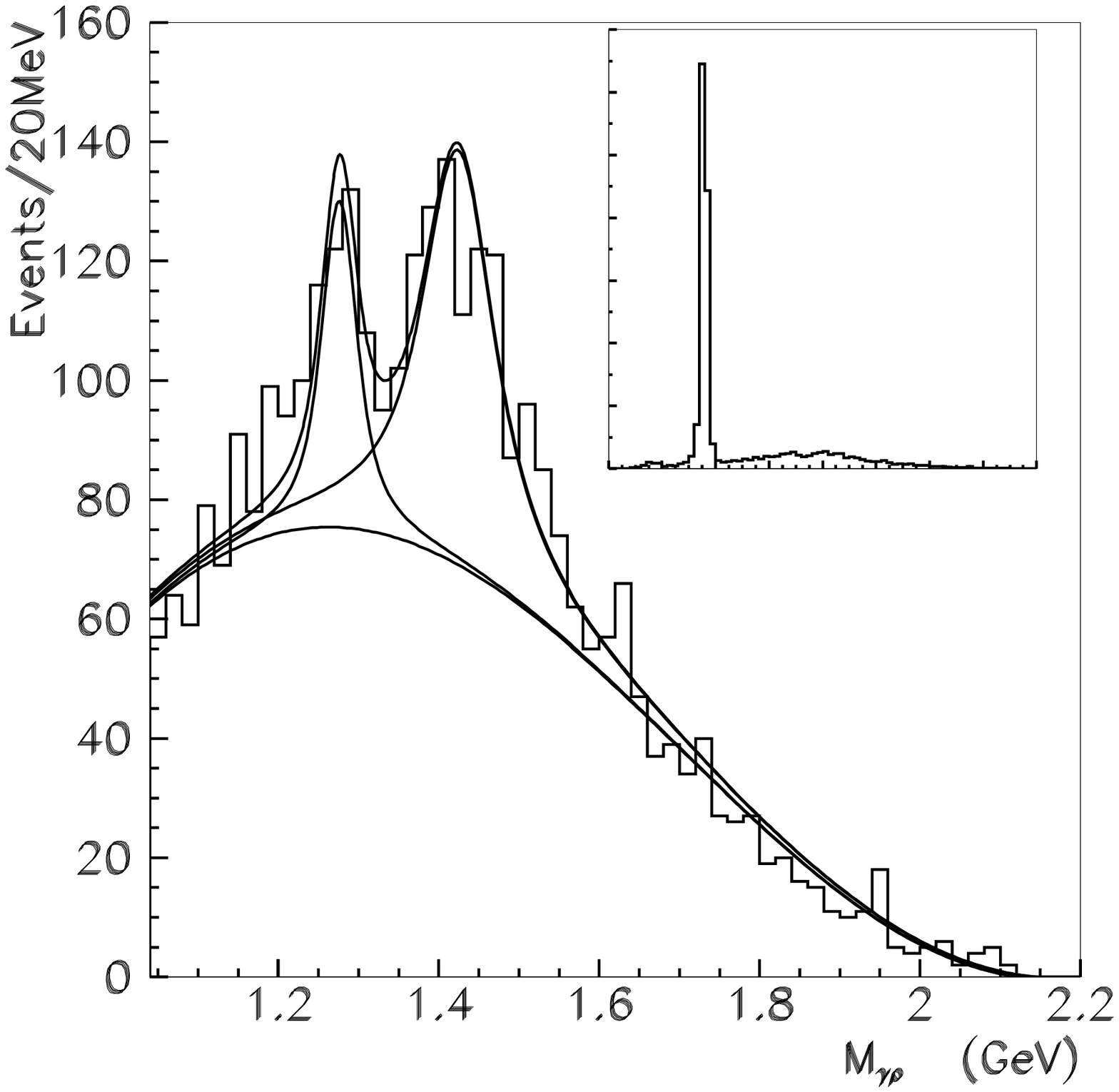,width=0.9\textwidth}}
    \caption{The $\gamma\rho$ invariant mass distribution. The insert
shows the full mass scale where the $\eta(958)$ is clearly
observed.}
    \label{ggrho}
\end{minipage}
\hspace{\fill}
\begin{minipage}{65mm}
\centerline{
    \psfig{file=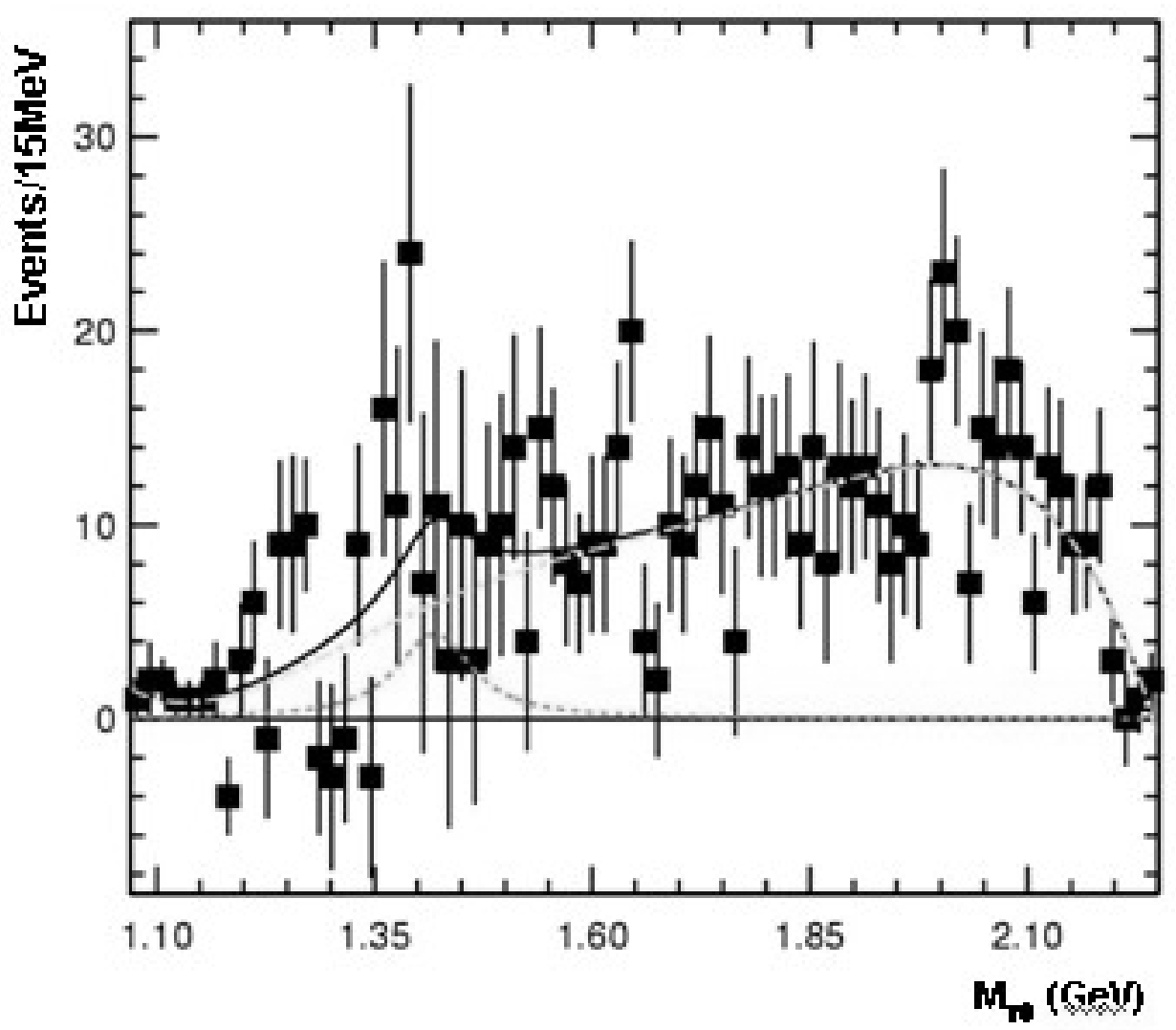,width=1.0\textwidth}}
  \caption{The invariant mass of $\gamma\phi$ after side-band
                        background subtraction.}
    \label{ggphi}
\end{minipage}
\end{figure}
\noindent
 In our $J/\psi\to \gamma\gamma V$ analysis~\cite{ggv},
there is a resonance around 1424 MeV at the $J/\psi\to \gamma\gamma
\rho$ channel. Comparing our result on the branching ratio
     $B(J/\psi \to \gamma X(1424) \to \gamma \gamma \rho) = (1.07 \pm
    0.17\pm 0.11 ) \times 10^{-4}$, and the
upper limit of B($J/\psi \to \gamma X(1424) \to \gamma \gamma \phi)
   <0.82 \times 10^{-4}$ (95\% C.L.), we cannot draw a definite conclusion
on wether the X(1424) is either a $q\bar{q}$ state or a glueball
state.

\par
\noindent
 We also analyzed the $\eta(1405)/\eta(1475)$ at $J/\psi\to
\{\omega, \phi\} K\bar{K}\pi$ channels~\cite{kkpi}. In the invariant
mass spectra of $K^{0}_{S}K^{\pm}\pi^{\mp}$ and $K^{+}K^{-}\pi^{0}$
recoiling against the $\omega$ signal region, the resonance at
$1.44$ GeV/$c^{2}$ is observed, while in the invariant mass spectra
of $K^{0}_{S}K^{\pm}\pi^{\mp}$ and $K^{+}K^{-}\pi^{0}$ recoiling
against the $\phi$ signal region, no significant structure near 1.44
GeV/c$^2$ is seen and an upper limits on the $J/\psi$ decay
branching fractions at the $90\%$ C.L. are given in
Table~\ref{kkpi}.
\begin{table}[bth]
  \centering
  \caption{The mass, width, and branching fractions of $J/\psi$ decays into ${\{\omega,\phi\}} X(1440)$.}
  \label{kkpi}
  \begin{tabular}{l|l}\hline\hline
$J/\psi \rightarrow \omega X(1440)$    & $J/\psi \rightarrow \omega X(1440)$  \\
($X\rightarrow K^{0}_{S}K^{+}\pi^{-}+c.c.$) & ($X\rightarrow
K^{+}K^{-}\pi^{0}$)  \\\hline
$M=1437.6\pm 3.2$ MeV/$c^{2}$    & $M=1445.9\pm 5.7 $ MeV/$c^{2}$       \\
$\Gamma=48.9 \pm 9.0$ MeV/$c^{2}$  & $\Gamma=34.2\pm18.5$
MeV/$c^{2}$ \\\hline \multicolumn{2}{l}{$B( J/\psi\rightarrow \omega
X(1440)\rightarrow \omega
K^{0}_{S}K^{+}\pi^{-}+c.c.)=(4.86\pm0.69\pm0.81) \times
10^{-4}$}\\\hline \multicolumn{2}{l}{$B( J/\psi\rightarrow \omega
X(1440) \rightarrow \omega K^{+}K^{-}\pi^{0})~~~~~~~~=
(1.92\pm0.57\pm0.38) \times 10^{-4}$} \\\hline
\multicolumn{2}{l}{$B(J/\psi\rightarrow \phi X(1440) \rightarrow
\phi K^{0}_{S}K^{+}\pi^{-}+c.c.)<1.93 \times 10^{-5}$} ($90\%$
C.L.)\\\hline \multicolumn{2}{l}{$B( J/\psi \rightarrow \phi X(1440)
\rightarrow \phi K^{+}K^{-}\pi^{0})~~~~~~~~< 1.71 \times 10^{-5}$}
($90\%$ C.L.)\\\hline \hline
\end{tabular}
\end{table}

\section{New Enhancements}
A narrow enhancement is observed in $J/\psi\to \gamma
p\bar{p}$~\cite{ppbar}. Assuming that the $p\bar{p}$ system is in an
S-wave resulted in a resonance with mass $M=1859^{+~3+~5}_{-10-25}$
~MeV/c$^2$, width $\Gamma<30$~MeV/c$^2$ (at the $90\%$ C.L.) and
product branching fraction $B(J/\psi\to\gamma X)\cdot B(X\to
p\bar{p})$ = $(7.0\pm 0.4 (stat)^{+1.9}_{-0.8}(syst)) \times
10^{-5}$. The data not precise enough to determine the angular
distribution. According to the theoretical calculation~\cite{yan},
if the $X$ is a bound state of $(p\bar{p})$, the decay channel
($X\to \eta 4\pi)$ is favored over $(X\to \eta 2\pi, 3\eta)$.
\par
The decay channel $J/\psi\to\gamma \pi^+\pi^-\eta^\prime$ is
analyzed using two $\eta^\prime$ decay modes, $\eta^\prime\to
\pi^+\pi^-\eta$ and $\eta^\prime\to\gamma\rho$~\cite{x1835}. A
resonance, the $X(1835)$, is observed with a high statistical
significance of 7.7$\sigma$ in the $\pi^+\pi^-\eta^\prime$ invariant
mass spectrum. From a fit with a Breit-Wigner function, the mass is
determined to be $M = 1833.7\pm6.1(stat)\pm2.7(syst)$~MeV/c$^2$, the
width is $\Gamma = 67.7\pm20.3(stat)\pm7.7(syst)$~MeV/c$^2$, and the
product branching fraction is $B(J/\psi\to \gamma X)\cdot B(X\to
\pi^+\pi^-\eta^\prime)$ = $(2.2\pm0.4(stat)\pm0.4(syst)) \times
10^{-4}$. The mass and width of the $X(1835)$ are not compatible
with any known meson resonance~\cite{pdg06}. If we redoing the
$S$-wave BW fit to the $p\bar{p}$ invariant mass
spectrum~\cite{ppbar} including the zero Isospin, $S$-wave
final-state-interactions (FSI) factor~\cite{fsi2}, yields a mass $M
= 1831 \pm 7$~MeV/c$^2$ and a width $\Gamma < 153$~MeV/c$^2$ (at the
90$\%$ C.L.), these values are in good agreement with the mass and
width of $X(1835)$ reported here.
\par
In the analysis of $J/\psi\to\omega p\bar{p}$~\cite{wpp}, no
significant enhancement near the $p\bar{p}$ mass threshold is
observed, and an upper limit of $B(J/\psi\to\omega X)B( X\to
p\bar{p})< 1.5 \times 10^{-5}$ is determined at the 95$\%$
confidence level.

\section{Summary}
Using the 58~M $J/\psi$ events sample taken with the BESII detector
at the BEPC storage ring, BES experiment provided many interesting
results, especially for the study of the lowest glueball candidates,
the structure of $\eta(1440)$, and the new enhancement of $X(1835)$,
but since the limit of the statistics, the better results (with
higher statistics and better accuracy) will be needed for well
understanding. The upgraded BEPCII/BESIII will provide a huge
$J/\psi$ decay samples for the further analysis.

\end{document}